# Applications of Novel Techniques to Health Foods, Medical and Agricultural Biotechnology

I.C. Baianu*, P.R. Lozano, V.I. Prisecaru and H.C. Lin

University of Illinois at Urbana, ACES College, email**: i-baianu@uiuc.edu**
FSHN Dept., Agricultural & Food Chemistry NMR & NIR Microspectroscopy Facility,
286 Bevier Hall, 905 S. Goodwin Ave, Urbana, IL. 61801, USA

## ABSTRACT

Selected applications of novel techniques in Agricultural Biotechnology, Health Food formulations and Medical Biotechnology are being reviewed with the aim of unraveling future developments and policy changes that are likely to open new markets for Biotechnology and prevent the shrinking or closing of existing ones. Amongst the selected novel techniques with applications in both Agricultural and Medical Biotechnology are: immobilized bacterial cells and enzymes, microencapsulation and liposome production, genetic manipulation of microorganisms, development of novel vaccines from plants, epigenomics of mammalian cells and organisms, and biocomputational tools for molecular modeling related to disease and Bioinformatics. Both fundamental and applied aspects of the emerging new techniques are being discussed in relation to their anticipated, marked impact on future markets and present policy changes that are needed for success in either Agricultural or Medical Biotechnology. The novel techniques are illustrated with figures that attempt to convey-- albeit in a simplified manner-- the most important features of representative and powerful tools that are currently being developed for both immediate and long-term applications in Agriculture, Health Food formulation and production, pharmaceuticals and Medicine. The research aspects are naturally emphasized in our review as they are key to further developments in Biotechnology; however, the course adopted for the implementation of biotechnological applications, and the policies associated with biotechnological applications in the market place, are clearly the determining factors for future Biotechnology successes in the world markets, be they pharmaceutical, medical or agricultural.

---

\* Corresponding Author.

**KEY WORDS:**

Applications in Agricultural and Medical Biotechnology;

immobilized bacterial cells and enzymes;

microencapsulation and liposome production;

genetic manipulation of microorganisms;

genetic therapy in medicine;

development of novel vaccines from plants;

epigenomics of mammalian cells and organisms;

biocomputational tools for molecular modeling related to disease;

Bioinformatics;

Suggested policy changes for future and immediate success of Biotechnology in the world markets;

Health Food Applications;

Novel Vaccines from Plants;

Medical Applications of Lectins for diagnostics and treatments;

Nuclear Magnetic Resonance (NMR) and MRI/ MRM;

Fluorescence Correlation Spectroscopy and applications to Single Molecule Detection, Single Virus Particle Detection;

Single virus particle, HIV, HPV, Hepatitis B and C, detection for early, successful treatments;

Ultra-sensitive, Selective Detection and Early Diagnostics of Cancers.

# 1. INTRODUCTION

Biotechnology has been traditionally thought to be associated with biomedical, and especially, pharmaceutical applications. Recently, agricultural biotechnology promises to yield even greater and wider gains through the enhancement of crop productivity or the use of transgenic crops than those that medical biotechnology has already achieved by exploiting combinations of cellular and molecular biology techniques. Novel biotechnology research is also rapidly expanding with a view to the application of environment friendly bioprocesses for chemical, pharmaceutical, and other areas, including food bioengineering and safety. The tremendous growth of the two areas of modern biotechnology (i.e., agricultural and medical) may have recently created the impression of technological isolation through specialization of both agricultural and medical biotechnology. Such an impression-- if not checked from spreading-- might raise artificial barriers between these two important and closely related domains. Caused mainly by over-specialization, an increasingly 'narrow' approach to biotechnology might affect the evolution and sharing of existing biotechnological tools that were initially developed for different types of applications. For example, the current adoption of certain genetically modified foods and the administration of vaccines based on genetically modified organisms (GMO' s) in developing countries, EU and Japan raise distribution problems for which adequate solutions have yet to be found. Potential loss of certain crop markets (such as corn) that may have included 'GMO-mixed' crops could be very substantial (>1 billion US $). Therefore, solutions to such problems must, and can be, found both through crop growing policy changes and also by utilizing novel, ultra-sensitive, as well as less expensive, techniques for GMO detection and crop quality control.

Unfortunately, about one billion people in the world remain deficient in several vitamins and minerals intake (Toenniessen et al 2003). Such undernourished populations live mostly in developing countries that often have inhospitable climates which favor low efficiency in food engineering and production, and also where the primary services for storage or administration of GMO-based vaccines are minimal. The utilization -- in combination with agronomical techniques-- of biotechnological tools that were originally produced for the medical area, may help resolve such difficult problems. As an example, the production of vaccines from plants and microorganism with subsequent addition in foods could reduce the cost of vaccination in such areas. The modification of microorganisms for the bio-production of fuels (Linko 1985) is also important, and would also be important in under-developed, or developing, countries that have a large surplus of under-utilized complex carbohydrate sources that could be used for bio-production of valuable fuels with existing technologies.

Our review focuses on novel techniques and tools employed in biotechnology that are being developed for chemical, medical, health food production and agricultural applications worldwide. However, no attempt or claim at extensive coverage of the subject is here being made.

## 2. IMMOBILIZED MICROBIAL CELLS

Immobilization of whole cells has been defined as the physical confinement or localization of intact cells to a certain defined region of space with preservation of some (Karel et al 1985), or most, catalytic activity. The increased stability under extreme conditions of pH and temperature, as well as the re-use and applicability in continuous processing systems that enclose immobilized cells instead of soluble enzymes make the cells a preferred, versatile tool in both food industry and medicine. There are several different approaches to the classification of immobilized biocatalysts, but the most frequently employed classification is based upon the method of immobilization selected for a specific application. The selection of immobilization method depends, therefore, upon the application, the nature of the microorganism being immobilized, as well as the resources available (Witter 1996). Table 1 shows several, possible immobilization methods that available for whole microbial cells. Most immobilization methods can be applied either to whole cells or to enzymes. Some of the advantages of whole-cell immobilization in comparison with enzyme immobilization are: the higher stability and enzyme activity, multivariate enzyme applications, and the lower cost (Witter 1996). On the other hand, disadvantages of using whole cell immobilization in comparison with enzyme immobilization are linked to the increased diffusional barriers caused by the much larger sizes of cells in comparison with enzymes.

| Binding | Weak bonds | Flocculation Adsorption Ionic |
| | Strong bonds | Covalent Cross-linking |
| Physical retention | Entrapment | Thermal Gelation Ionotropic Gelation Polymerization |
| | Membrane Retention | Dialysis Culture Ultrafilters |

Table 1. Immobilization Methods for Whole Microbial Cells.
(SOURCE: Witter 1996, Ch. 10 in: " Physical Chemistry of Foods" Vol.2, Baianu et al Eds. 1996).

Adsorption is the least expensive and mildest immobilization method. It uses weak interaction forces such as hydrogen bonds, hydrophobic interactions and Van der Waal forces to immobilize cells or enzymes. However, the sensitivity of this interaction to pH makes the leakage of cells immobilized by this technique quite common. Important applications of this technique are related to the production of fructose and vinegar, and also waste water treatments.

Ionic binding uses the properties of negatively charged microbial cells to interact with positively charged ion exchangers (Witter 1996). The results obtained with this technique are also sensitive to extreme pH values, and the binding strength is greater in comparison with adsorption. However, the mild conditions employed by this technique make it suitable for use for immobilization of both enzymes and whole cells.

Covalent binding and cross-linking offer better strength than the previous techniques, however, there is an encountered toxicity in the reagents (Chibata 1979; Kolot 1981; Babu and Panda 1991) that are used to produce immobilization.

Entrapment techniques are, however, the most commonly encountered in the industry and they are based on the formation of thermally reversible gels, ionotropic gels and polymerization.

## 2.1. Application of Cell Immobilization in the Food Industry and Human Nutrition

Entrapment of cells in a gel–like matrix by ionotropic gelation using alginates and K-carrageenans is certainly the most useful method for industrial purposes. The alginates form a gelatin matrix in the presence of polyvalent cations by binding the cation to guluronic acid units (Witter 1996; Nedovic et al 2003) whereas *k-* carrageenan solidifies in the presence of potassium ions. The properties of the gel-like matrix allow the cell to remain viable and with its catalytic ability for a long period of time. For example, an increment in the yeast concentration obtained through immobilization techniques has helped the brewing industry to reduce fermentation process times and the size of their storage facilities. Unfortunately, because of the high concentration of diacetyl, and the low concentration of higher alcohols and esters, the flavor of the fast fermented beer has been compromised. Even the amino acid profile has been altered (Masschelein et al 1994). The main factor causing this uncommon imbalance is the insufficient mass transfer in the older designs of fermentation reactors; thus, the use of new reactor designs (Andries et al 1996; Nevodic et al 1996), and combining technologies (Inoue 1995) could improve the quality of the products obtained through such fermentation reactions.

Nutraceuticals are defined as food components that have health benefits beyond traditional nutritional value. Novel biotechnology tools like immobilization were also applied for the isolation and incorporation of such food components in ordinary foods. The synthesis of nutraceuticals was reported to be successful by employing immobilized lipases, such as those from *Candida antartica* and *Lactobacillus ruteri* (Hill et al 1999, and Garcia 2001). The introduction of conjugated linoleic acid (CLA) in dairy foods has been made possible through the immobilization of lipases (Hill et al 1999).

### 2.2. Medical Applications of Molecular and Cellular Immobilization Techniques

There is a quite extensive list of immobilization technique applications in medicine.
A very important group of such applications is concerned with the regulation of equilibrium between coagulation and dissolution of coagulated blood (fibrinolysis) through the use of immobilized enzymes. The high probability of death caused by thrombosis (involving the formation of clots in the blood vessels), has committed physicians to the use of fibrinolytic therapy for the treatment of occlusions in those parts of the body where a surgical intervention would be too risky. Amongst the most important enzymes that have been immobilized for use in such therapy are Plasmin and Heparin (Wolf and Ransberger 1972). The use of biotechnology as well as microscopic techniques has helped refine and greatly improve such therapeutical means.

### 2.3. Other Applications of Novel Immobilization Techniques

Current regulations for the disposal of toxic chemicals in the environment as well as the detoxification of water used in any agricultural and industrial process brings the need for novel biotechnology tools to be developed in order to solve such problems in a cost-efficient manner. Enzymes have been isolated from genetically manipulated microorganism strains with the purpose of accelerating the rate of degradation of organic and some inorganic compounds in wastewater as well as in soils (Head and Bailey 2003).

## **3.** GENETIC MANIPULATION OF MICROORGANISMS FOR BIOTECHNOLOGY APPLICATIONS

Genetic manipulation techniques have played an important role in enhancing the performance of microorganisms that are significant in the industrial, pharmaceutical and medical application areas. Such techniques are discussed briefly in this section together with their selected applications. It is important to highlight the fact that they are generic, and also that they can be employed to improve the performance of any cell or microorganism for specific applications.

### *3.1. Mutagenesis-based Techniques*

*Mutagenesis* can be defined as the group of techniques or processes that produce changes in a DNA sequence which modifies either the expression of genes or the structure of the gene products. Two subcategories of mutagenesis are usually applied in the genetic manipulation of microorganisms: classical mutagenesis and transposon-directed mutagenesis.

*Classical Mutagenesis* involves the use of chemical mutagens to regulate a process of interest in the target microorganism. The main purpose of this 'classical' approach is to find the mechanism of action of the targeted microorganism.

*Transposon-directed Mutagenesis* consists in the genetic alteration of a few sites on the target chromosome in order to determine the function of this specific site of the chromosome. By comparison with the previous category transponson- directed mutagenesis it is more specific, and is therefore preferred in specific applications that require higher selectivity.

### *3.2. Gene Transfer*

*Electroporation* involves the production of transient pores in a bacterial cell membrane following the application of high-voltage electric field (DC) pulses of short duration . Such pores allow the introduction of DNA into the cell under certain, favorable conditions. Such methods have been used with various microorganisms and even plasmids that are of significant use in the industry.

Another technique that is employed for the transference of genes involves Shuttle Vectors. *Shuttle vectors* are DNA constructs that are able to replicate and deliver DNA to widely divergent types of bacteria (Blaschek 1996). DNA transference is obtained after the use of a microorganism whose mechanisms of action have been thoroughly studied, as for example, is the case of various *E. coli* strains.

### 3.3. Gene Cloning

Gene cloning and genomic maps to clone the genes of industrially-significant microorganisms have also been reported to be subsequently introduced in native strains of the microorganism of interest (Zappe et al 1986; Cary et al 1990; Verhasselt 1989).

### 3.4. DNA Analysis

DNA analysis involves the study of a particular region of DNA by either physical or chemical techniques. It is usually carried out in order to obtain insights into the genetic organization, the mechanisms of gene expression , or to obtain information that may be useful for the construction of shuttle vectors (Blaschek 1996)

### 3.5. Applications in the Food Industry

Pioneering studies of genetic manipulation of microorganisms with significant applications to foods were initially based on *Escherichia coli* as a microbial model because of its ease of growing and manageability. Subsequent studies were carried out in the production of dairy starters, as well as the amino acid and yeast production for brewing purposes, and more recently attempts are being made to produce butanol through fermentation.

## 4. EPIGENOMICS IN MAMMALIAN CELLS AND ORGANISMS

Upon completion of the US Human Genome Mapping Project and related studies, it became increasingly evident that a sequence of 30,000 or so 'active' genes that encode and direct the biosynthesis of specific proteins could not possibly exhaust the control mechanisms present in either normal or abnormal cells (such as, for example, cancer cells). This is even more obvious in the case of developing embryos or regenerating organs. Recently, specific control mechanisms of cellular phenotypes and processes were proposed that involve *epigenetic* controls, such as the specific acetylation<—> de-acetylation reactions of DNA-bound histones (see, for example, *Scientific American* 2003, December issue, for a recent review article on epigenomics).

Epigenomic tools and novel techniques begin to address the complex and varied needs of epigenetic studies and their applications to controlling cell division and growth. Such tools are, therefore, potentially very important in medical areas such as cancer research and therapy, as well as for improving 'domestic' animal phenotypes _without_ involving genomic modifications of the organism. This raises the interesting question if 'epigenomically controlled-growth organisms (ECGOs) -- to be produced in the future-- would be still objected against by the same groups of people that currently object to GMOs , even though genetic modifications would be neither present nor traceable in such organisms ?

## 5. MICROENCAPSULATION TECHNIQUES AND THEIR APPLICATIONS

Microencapsulation is now being extensively used in the industry and medicine to enhance the stability and shelf life of products, to improve the nutritional value of foods, and to produce enzyme, or immuno-protective, capsules for the production of vaccines and for the transplant of organs. However, the prospects of microencapsulation techniques are not limited to these areas because it can be used in combination with other novel biotechnology tools to broaden the spectrum of its applications.

The concept behind *microencapsulation* involves surrounding a component of interest (such as vitamins, proteins, flavor, etc.) that need protecting with a (μm-thick) thin layer of either polysaccharide or lipid material in order to prevent, or significantly reduce, the deterioration of the encapsulated component. As a result, liquid droplets, solid particles or gaseous materials are packaged with continuous shells that can release their contents at controlled rates under desired conditions (Rosenberg and Kopelman 1983).

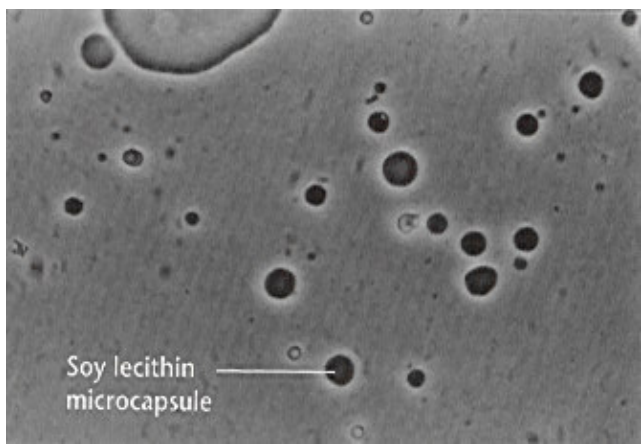

Fig 1. Photomicrograph of Soy Lecithin microcapsules of about one micron in size.
(SOURCE: Baianu et al 1993).

There are several, 'classical' techniques that are used for microencapsulation such as: spray drying, molecular inclusion and extrusion. On the other hand, liposome microencapsulation is a novel technique that is being developed to the point that it can introduce any substance of interest in an organism regardless of the substance's solubility, electrical charge, molecular size or structure (Gregoriadis 1984 ).

### 5.1. *Liposomes and Liposome Microencapsulation*

A *liposome* can be defined as an artificial lipid vesicle that has a bilayer phospholipid arrangement with the head groups oriented towards the interior of the bilayer and the acyl group towards the exterior of the membrane facing water (**Fig.1**). Liposomes are usually made of phosphatidylcholine (lipid) molecules although mixtures of phospholipids can also be employed to make liposomes. The main difference between liposomes and cell membranes is the lack of bilayer asymmetry in the case of liposomes. On the other hand, liposomes do exhibit barrier properties very similar to those of cellular lipid membranes, and therefore, they are frequently used as a model system for the study of the transport activity of the cell membrane boundary (Smith 1996).

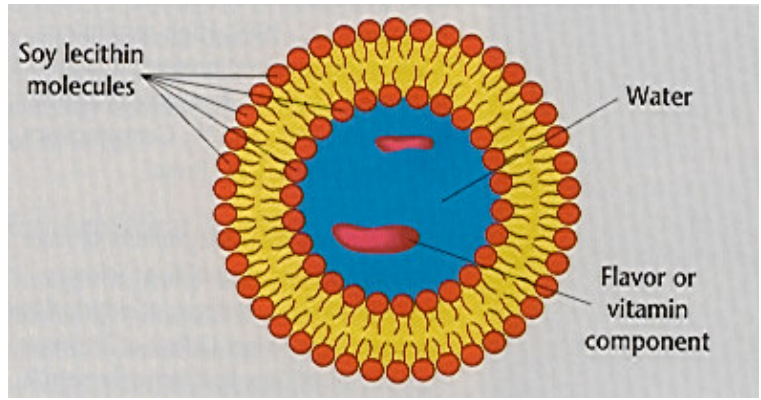

Fig 2. Simplified representation of the molecular organization of a liposome microcapsule in water.

Since the invention of liposomes in the early 1960's as 'membrane-like' structures, they have been extensively used in the biotechnology field. The 'membrane-like' characteristics, their colloidal size (from ~30 nm to 10 μm) as well as their solubility properties for various molecules have made liposomes useful tools for various biotechnology applications, from foods applications (such as in enzyme microencapsulation), to fermentation and several medical/ pharmaceutical applications.

A few possible mechanisms for liposome absorption into cells are illustrated in Fig 3. Fusion and endocytosis are the most common ones being utilized for the production of liposomes with industrial or medical purposes. During *fusion*, the soluble material carried in the liposome is released into the cytoplasm of the target cell through the joining of the liposome with the cell membrane lipids. Such a mechanism is useful for introducing hydrophobic material into the cell, and it has been already employed by the petroleum industry (Hofe et al 1986).

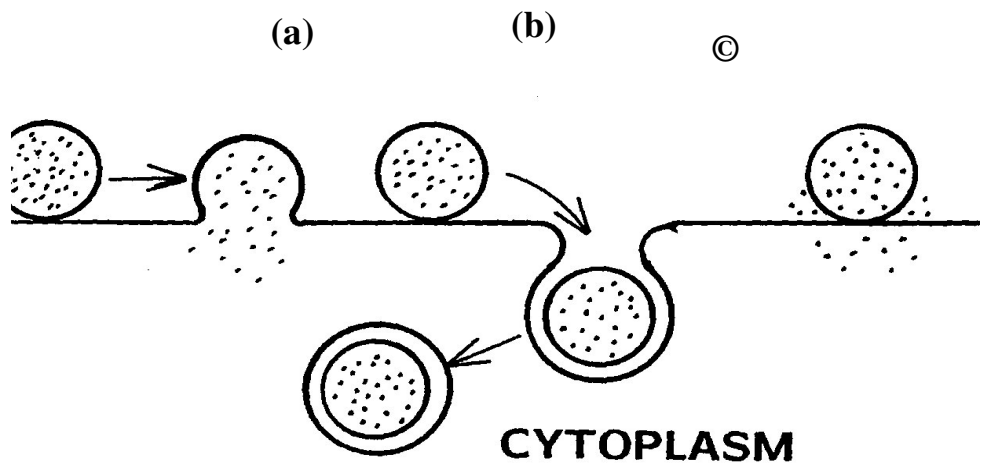

Fig 3. Possible interactions of a liposome with a cell membrane: (a) Fusion;  (b) Endocytosis;
(c) Adsorption. (SOURCE: Ch .11 in *"Physical Chemistry of Food Processes"*, vol 2,
Baianu  et al, eds. 1996).

In the case of *endocytosis*, the liposome enters the cell surrounded by an endocytotic vacuole that is derived from the membrane lipids. The contents of the liposomes are then released by the action of liposomes that were attached to the produced vacuole.  Low molecular weight compounds (that are not charged at low pH values) will thus be able to diffuse into the cytoplasm.

### 5.1.1. Classification and Production of Liposomes

Liposomes are classified according to their size and structure, the latter also depends on the preparation method (Deamer and Uster 1983; Gregoriadis 1984)**.** There are three classes of liposomes: multilamellar vesicles (MLV's), small unilamellar vesicles (SUV' s) and unilamellar vesicles (LUV's; Gregoriadis 1984). The type of liposome being utilized in a specific application depends on the hydrophobicity of the molecule being encapsulated.

MLV's consist of a series of multiple lipid bilayers that are obtained from a phospolipid solution in chloroform, which is then evaporated to produce a thin film and subsequently hydrated in an aqueous solution to form heterogeneous vesicles with diameter sizes from ~0.3 to 2.0 μm. The main advantage of the MLV's is that both the lipids and the aqueous solution to be encapsulated are not subjected to harsh treatment (Kim and Baianu 1991).

The use of high-intensity ultrasonication, ethanol injection and pressure applied to multillamelar vesicles allows the production of single bilayer vesicles SUV's (Deamer and Uster 1983); this physical treatment reducesthe liposome size. Unfortunately, the smaller size of such a vesicle also results in a higher curvature that limits its capture volume.

LUV's are prepared commonly by infusion, reverse–phase evaporation and detergent dilution methods. They vary in size range from 100 -500 nm and they are the most widely employed liposomes (Kim and Baianu 1991) because they have less variable size and higher entrapment volumes (>~2.7 L / mol of lipid ) than SUV's (Smith 1996) .

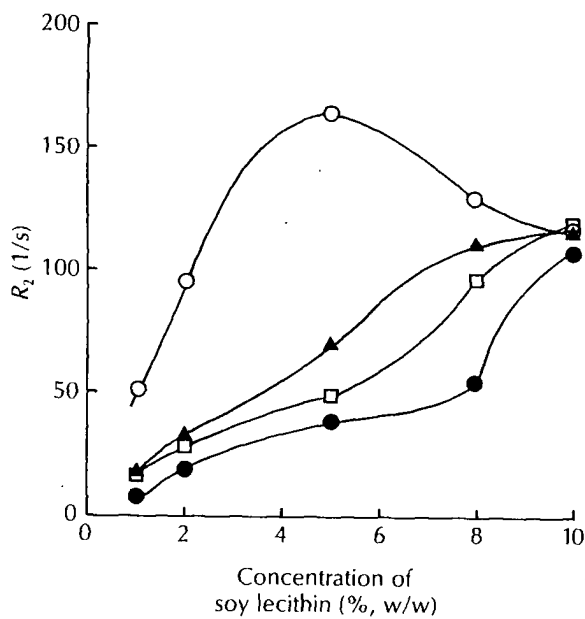

**Fig 4**. The effect of Soy Lecithin Concentration on the Storage Life of Liposomes. Release of $Mn^{+2}$ during storage is indicated by the marked increase of the relaxation rates (**T$_1$** and **T$_2$**) measured by NMR relaxation techniques:  (●)immediately after $Mn^{+2}$  microencapsulation, (□) after one day of storage , ( ▲) after two days ,  and finally (○) after treatment with the Triton X –100 detergent. (SOURCE:  Kim and Baianu 1993).

### 5.1.2. Liposome Formation

The phase diagram in **Figure 5** illustrates the phase changes around the transition temperature ($T_c$) of a phospholipid- water system. This phase transition temperature is defined as the minimum temperature required for the water to break through the lipid membrane. When the system is cooled to temperatures below $T_c$, the hydrocarbon chains adopt an ordered packing phase, thus creating a lamellar structure (Chapman et al. 1967). In the bilayer structure, the hydrophobic tails are lined up together through hydrophobic interactions, whereas the hydrophilic part of the lipid faces towards the aqueous phase.

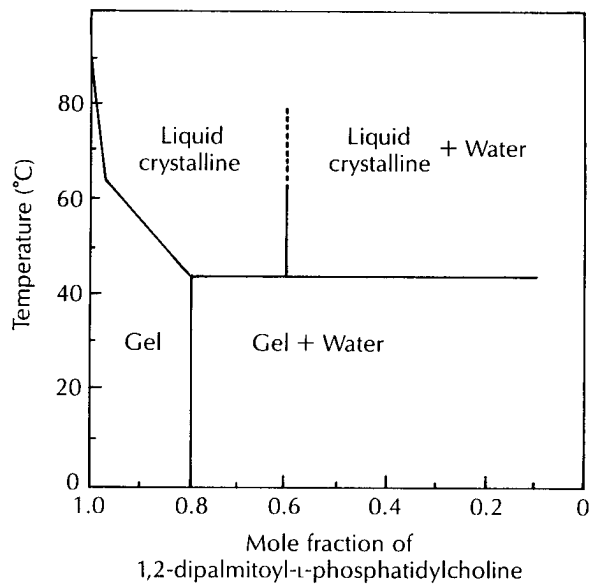

Fig 5. Phase diagram of 1,2-dipalmitoyl-l- phosphatidilcholine water. From Chapman et al 1967.

### 5.1.3.    Techniques for Liposome microencapsulation

The lipid-aqueous system needs to meet two major requirements for the liposome microencapsulation to occur. First, the system needs have a negative free energy, and second, it needs to overcome the energy barrier which is necessary for the formation of the bilayer. Three methods of liposome preparation are here described:

### A.  Microfluidization

Microencapsulation by this method is obtained through the dynamic interaction of two pressurized aqueous-lipid fluids that create a large momentum and flow turbulence that allows the system to overcome the energy barrier to microcapsule formation. The pressure applied in air-driven microfluidizers can be as high as 10,000 psi (Kim and Baianu, 1993). The ultra-high velocities reached by this technique allow the creation of small liposomes (< ~0.3μm) with high capture efficiency (Mayhew and Lazo 1984). This system is useful because of its capability to produce very large amounts of liposomes with adjusted size in a continuous process.

## B. Ultrasonication

The ultrasound absorption is employed to overcome the energy barrier. The sonication of the lipid emulsion can be carried out by two different approaches. The first one is through the use of a sonication probe placed directly into the suspension of liposomes. The second method is slower than the first one and employs a sonication bath, such as a sealed container filled with Nitrogen or Argon gas. Both methods have been extensively applied for the formation of SUV' s; however, the use of a sonication probe has been found to cause contamination of the liposomes (Taylor 1983).

## C. Reverse phase evaporation

Reverse phase evaporation is used for the preparation of LUV's (Szoka and Papahadjopoulos 1978) and it is based upon the extraction of a nonpolar solvent from an aqueous–nonpolar inverted micelle by rotatory evaporation under vacuum. This withdrawal of the nonpolar phase changes the intermediate gel-like phase of the micelle into uni-lamellar and oligo-lamellar vesicles (Kim and Bainu 1993). The advantage of this technique is the uniformity of the vesicles formed (from about 0.2 to1.0 μm) as well as their high encapsulation efficiency. On the other hand, the exposure of the components to organic solvents and sonication is likely to result in protein denaturation (Szoka and Papahadjopoulos 1980)**.**

## 5.1.4. Characterization of Liposomes

Several techniques may be employed to characterize liposomes. **Table 3** compares some of the current techniques and novel approaches to the study of liposomes and their interactions.

| Technique | Liposome Characteristics and Related Processes |
|---|---|
| *Electron Microscopy* | Liposome Size distribution<br>Average size of Liposomes |
| *Radiactive Tracers* | Liposome stability vsus time and temperature<br>Encapsulated Molecule Retention rates (Mayhew et al. 1983) |
| *Fluorescence Quenching* | Liposome stability<br>Liposome-cell interactions |
| *Ultrasonic Absorption* | Temperature induced transitions<br>Limitations of Liposome uses in the presence of hydrophobic proteins |
| *Electron Spin Spectroscopy* | Permeability of liposomes<br>Molecular dynamics of Spin- Labeled Lipids in Liposomes and Encapsulates |
| *Nuclear Magnetic Resonance* | Molecular dynamics of Lipids in Liposomes and Encapsulates<br>Water exchange in phospholipid vesicles (Haran and Sporer 1976)<br>Stability using metal cations (Baianu and Kim 1993)<br>Storage of Liposomes |
| Fluorescence Correlation Spectroscopy | DNA interactions with molecules encapsulated within liposomes<br>Hybridization of DNA for Liposomes Vaccines |

Table 3. Characterization of Liposome Properties by Various Techniques.

### 5.1.5.     *Applications of Liposomes in the Food Industry*

Lecithin-based liposomes offer great flexibility and shelf life improvements in the food industry for introducing water–soluble substances such as flavors and micronutrients. Such modified liposomes are also being used at present in the beverage and bakery industries to deliver both flavors and oils that are trapped inside the liposomes; such flavors and oils are released then into the mouth; these are also employed to incorporate flavor oils. The major impact of such techniques has been achieved through the use of microencapsules that can be made through a continuous process on an industrial scale (**Fig 6**).

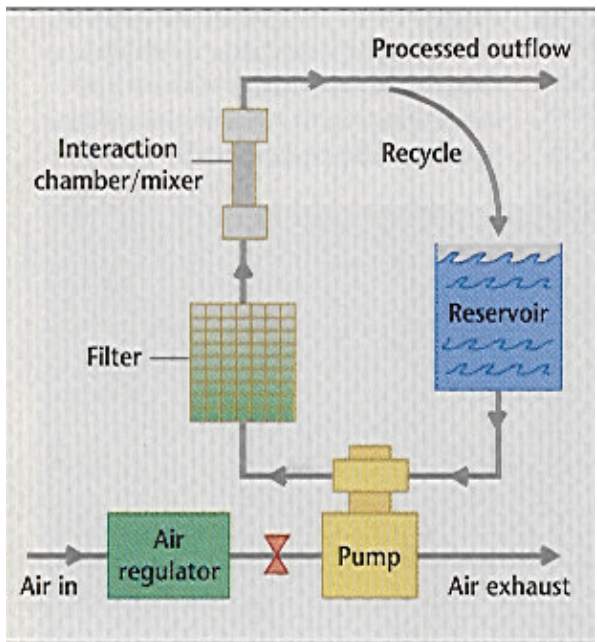

**Fig 6.** Microfluidizer employed to microencapsulate enzymes and food proteins in food products.

In the dairy industry, liposomes containing enzymes have been reported to reduce the ripening time by 30 -50% (Kirby 1990 and Law and King 1991), as well as improve texture and flavor. The latter was caused by a decrease in the action of proteolytic enzymes in the early phases of cheese fermentation. (Alkhalaf et al 1988).

Because liposomes have the ability to carry fat-based flavors in their bilayer, as well as water-soluble flavors in the core of the vesicle, they protect the flavor from degradation and also increase the longevity of the flavor in the system where they are being employed. Therefore, their use in the beverage industry has become widespread. The rate of diffusion through/from the bilayer depends on the liposome composition as well as physical properties of the flavor.

Bakery is another area where liposomes have been applied and it is based on the characteristic of the liposomes of not being destroyed during the processes of mixing or extrusion; therefore, they can release encapsulated flavorings, fragrances or food additives. When a flavor is encapsulated, the release occurs after the enzymatic degradation of the liposome, and thus the rate of release depends on the physical properties of the material of which the liposome is made. In the case of lecithin, the pH value or range, as well as temperature, are important factors.

### 5.1.6. *Liposome Applications in Medical Biotechnology*

Traditionally, liposomes have been used in the bioengineering field to over-produce certain proteins through the genetic modification of cells. They have thus solved the inconvenience of transferring high molecular weight molecules through cell membranes (Nicolau and Cudd 1989).

The use of unilamelar vesicles made of cationic lipids (Rose et al 1991) has improved the transfection efficiencies and prevented interactions with DNA molecules. Therefore, the project of introducing genes to cure diseases (*genetic therapy*) is not far from becoming a reality if the patient being treated were found not to suffer from severe side effects. Vaccine formulations based on liposomes have been successfully tested in animal immunization, and such studies are currently in the clinical testing phase. The benefits and limitations of liposomes as drug carriers in a system as complex as the human body depend basically on the interactions of liposomes with the cells (Lasic, 1993), as well as their immuno-compatibility, or their ability to escape detection by the human immune system.

### 5.1.7. *Other Applications of Liposomes*

Cosmetics are another area where the liposomes have been extensively employed. They are being utilized as humectants, as well as carriers of formulations containing extracts, vitamins, moisturizers, antibiotics and proteins. Such applications are mostly directed towards preventing, or delaying, the aging of the skin. Through their surfactant action liposomes also improve the coagulation and sinking of oil spreading at a water interface, a methodology which has been under study for some time by EPA for cleaning up oil spills (Gatt et al 1991; Dutton 1993).

## 6. LECTIN APPLICATIONS TO CANCER DETECTION AND TREATMENT

Lectins are proteins, or glycoproteins, that agglutinate erythrocytes of some or all blood groups *in vitro* (Sharon 1998). They are an important group of bioactive proteins and glycoproteins found in most organisms, including plants, vertebrates, invertebrates, bacteria and viruses, and have several important applications to the fields of health food and medical biotechnology. Lectins are used as tools in the fields of biochemistry, cell biology and immunology, as well as for diagnostic and therapeutic purposes in cancer research (Sharon and Lis 2002). Lectin aggregation can be employed on a large-scale basis for the commercial production of biologically active proteins (Takamatsu et al 1999).

Lectins have been used in glycoprotein purification, oligosaccharide analysis, as well as in cell-selection processes. Lectins can bind reversibly with free sugars or with sugar residues of polysaccharides, glycoproteins or glycolipids (Goldstein and Poretz 1986). There has been increasing demand for novel diagnostic and medical cancer therapies that utilize non-traditional sources. Epidemiological studies indicate that the consumption of a plant-based diet is strongly associated with a reduced risk of developing several types of cancer (Block et al 1992). Plants contain numerous phytochemicals that can alter cancer-associated biochemical pathways. One such group being intensively examined for its role in cancer chemoprevention is lectins (Abdullaev and Gonzalez de Mejia 1997). A review of plant lectins and anticancer properties can be found in a review article (de Mejia and Prisecaru 2003), and soybean lectins are specifically discussed in another publication (Gonzalez de Mejia et al 2003).

Lectins are currently being considered for use as cancer therapeutic agents. Lectins were reported to bind preferentially to cancer cell membranes, or their receptors, causing cytotoxicity, apoptosis and/or tumor growth inhibition. Lectins were thought to become internalized into cells, and some lectins were claimed to cause cancer cell agglutination and/or aggregation. Present in common foods, some lectins resist acid and/or enzymatic digestion and also were reported to enter the bloodstream in an intact, and biologically active, form. Lectins possess a spectrum of beneficial, as well as harmful, effects both *in vitro* and *in vivo*. Ingestion of lectins also sequesters the available body pool of polyamines, thereby claimed to thwart cancer cell growth. They have also been reported to affect the immune system by altering the production of various interleukins, or by activating certain protein kinases. Lectins were also reported to bind to ribosomes and thus inhibit protein synthesis. Lectins may also modify the cell cycle by inducing non-apoptotic G1-phase accumulation mechanisms, G2/M phase cell cycle arrest, and apoptosis, and might activate the caspase cascade. Lectins were also reported to down-regulate telomerase activity and inhibit angiogenesis. Lectins could inhibit cell adhesion, proliferation, colony formation and hemagglutination, and were reported to have cytotoxic effects on human tumor cells. Lectins could function as surface markers for tumor cell recognition, cell adhesion, signal transduction across the membrane, mitogenic cytotoxicity and apoptosis. Also, lectins were reported to modulate the growth, proliferation and apoptosis of premalignant and malignant cells both *in vitro* and *in vivo*. Most of these effects are thought to be mediated by specific cell surface receptors (Gonzalez de Mejia and Prisecaru 2003).

For many years, lectins have been considered toxic substances to both cells and animals, mainly because of the observed agglutination of erythrocytes and other cells *in vitro*. On the other hand, it has also been reported that lectins have an inhibitory effect on the growth of tumors. Their potential for clinical applications has been investigated only in recent years. Lectins are now being considered for use both in the diagnostics and therapeutics of cancers. Thus, lectins are quite versatile biomarkers and have been utilized in a variety of studies involving histochemical, biochemical and functional techniques for cancer cell characterization (Munoz *et al* 2001). Lectins may also be very useful tools for the identification of cancers and the degree of metastasis, or cancer development stage. Recently, there has been a tendency to shift lectins use from cancer detection to actual use in combating cancer. The reason for this shift is mainly caused by recent research that indicated the cytotoxic and apoptosis/necrosis-inducing effects of certain lectins, combined with the hypothesis that dietary lectins enter the systemic circulation intact (Wang et al 1998).

One important feature appears to be that lectins stimulate the human immune system. Lectins were thus reported to exhibit anti-tumor and anti-carcinogenic activities that could be of substantial benefit in cancer treatment. Extracts of *Viscum album* (mistletoe) are widely used as complementary cancer therapies in Europe. Mistletoe has been used parenterally for more than 80 years as an anticancer agent with strong immuno-modulating action. The quality of life of patients with late-stage pancreatic cancer was reported to be improved on account of exposure to mistletoe lectin (Friess et al 1996). Immuno-modulation using recombinant ML was reported to influence tumor growth in breast cancer patients (Stein et al 1998). Bladder carcinoma was reported to be significantly reduced, and survival times were reported to be prolonged in mice as the concentration of ML was increased from 3 to 30 ng. ML increased the life span, decreased the tumor growth and decreased hyperplasia of mice and rats with lymphoma and lung cancer (Kuttan et al 1997).

A lectin purified from mesquite seed was reported to have an anti-proliferative effect on the cervical human tumor (HeLa) cells and on cell adhesion. Interestingly, mesquite lectin modulated the growth, proliferation and apoptosis of HeLa cells, while having no effect on normal cells *in vitro* (Gonzalez de Mejia et al 2002; Abdullaev and Gonzalez de Mejia, 1996). *Vicia faba* agglutinin (VFA), a lectin from broad beans was reported to aggregate, stimulate the morphological differentiation of, and reduce the malignant phenotype of colon cancer cells (Jordinson et al 1999). Wheat germ agglutinin (WGA) proved to be highly toxic to human pancreatic carcinoma cells *in vitro*. WGA exposure induced chromatin condensation, nuclear fragmentation and DNA release, consistent with apoptosis. The binding of the snail lectin Helix pomatia agglutinin (HPA), which recognizes N-acetylgalactosamine and N-acetylglucosamine sugars, is considered to be a strong predictor of metastasis and unfavorable prognosis in a number of human adenocarcinomas, including breast cancer (Brooks and Carter 2001). Because of their carbohydrate bio-recognition properties, lectins may also be used as carriers for targeted drug delivery, in a manner similar to liposomes (Wroblewski et al 2001), provided the possible side effects of such treatments could be minimized.

It has been observed that mucosal expression of terminal unsubstituted galactose is increased in colon pre-cancerous conditions and cancer, and that it allows interaction with mitogenic galactose-binding lectins of dietary or microbial origin. Based on this observation, an interesting hypothesis was postulated whereby galactose might be able to prevent cancer by binding and inhibiting such lectins from interacting with colon cancer cells (Evans et al 2002). D-galactose treatment was reported to be effective in liver lectin blocking to prevent hepatic metastases in colorectal carcinoma patients (Isenberg et al 1997). Epithelial cancer cells showed an increased cell surface expression of mucin antigens with aberrant O-glycosylation, notably Thomsen-Friedenreich Antigens (TFA). TFA is a carbohydrate antigen with a proven link to malignancy (Irazoqui et al 2001). Immunoassays could be utilized for antigens such as TFA in order to determine the metastatic potential of breast and colon cancer cells. Molecular changes in the membrane surface in the case of both stomach and colon cancer cells occur during the progression to carcinogenesis. Carbohydrate patterns displayed on the cellular membrane exterior are *molecular signatures* with *unique* biological characteristics related to oncogenesis and metastasis, and could be used to determine the appropriate chemotherapeutic and surgical procedures for each specific cancer.

Lectins have already a demonstrated potential for the treatment, prevention and diagnosis of chronic diseases such as cancer. Further research is, however, required to further elucidate the effects of purified and dietary lectins and their potential for defense against tumors.

## 7. COMPUTATIONAL BIOLOGY, MOLECULAR MODELING AND SOME OF THEIR BIOTECHNOLOGY RELATED APPLICATIONS

Computational Biology has a very wide range of applications currently thought of 'belonging' to Biotechnology (Baianu 1986), such as Bioinformatics, even though the development of such computations has preceded modern Biotechnology by many decades. Instead of attempting the hopeless task of covering superficially just a few of the applications of Computational Biology to either Medical or Agricultural Biotechnology, we decided in favor of an approach that focuses on a few selected examples in greater depth, by considering molecular modeling techniques that have very wide applications, ranging from 'pure' chemistry, to biochemistry, molecular biology, biotechnology, medicine, foods and industrial manufacturing.

### 7.1. Molecular Modeling Techniques

*Molecular modeling* is a group of techniques that employ computer-generated images of chemical structures that show the relative positioning of all the atoms present in the molecule being studied, and/or the simulated dynamics of such molecules together with their ordering through space-time. Such techniques are of considerable help for understanding many physicochemical properties of molecules, and may also provide clues about their possible role(s), that is, their *function*, in the organism. They can be thus especially valuable tools for investigating *structure-function* relationships. Proteins --within a given protein family-- have, in theory, similar sequences and generally share the same basic structure. Thus, once the structure for one member of the protein family is determined, molecular modeling computations can help determine the structure for other members of the same protein family. Such a homology technique when applied to protein structure may allow scientists to gain additional insight into protein structure, especially for those proteins for which the available experimental data is scarce.

### 7.1.1 Tasks in Molecular Modeling

In order to obtain optimal results, the National Center by Biotechnology (NCBI) suggested that protein sequences should be organized in protein families. Such readily searchable databases (Table 4) are currently available for many proteins (NCBI 2003).

| Table 4. Databases for Molecular modeling |
|---|
| Protein Data Bank (PDB); |
| The molecular modeling database (MMDB) at NCBI |
| Clusters of Orthologous Group of proteins (COGs) with COGNITOR program |
| The Basic Local Alignment Search Tool (BLAST) |
| Vector Alignment Search Tools (VAST) |
| The Conserved Domain Database(CDD) |
| Domain Architecture Retrieval Tool. |

(Source:  NCBI 2003: "A science primer" ).

Secondly, a *target* must be selected. A *target* is a protein structure that has been determined *via* experiments. Thirdly, one must generate a purified protein for analysis of the chosen target and then determine the protein structure by analytical techniques such as X-ray crystallography and/or 2D NMR.

The experiments described next, for example, studied apolipoprotein A-1 by employing molecular modeling techniques in order to understand the interaction of proteins in food systems and complex organisms.

### 7.1.2. Apolipoprotein Structures

Lipoprotein in mammals have evolved as the primary transport vehicles for lipids. This role leads to the importance of lipoproteins in several diseases, such as atherosclerosis and cardiovascular disease. Lipoprotein particles consist of a core of neutral lipids, stabilized by a surface monolayer of polar lipids complexed with one or more proteins.

Apolipoprotein A-I and apo B are respectively, the major protein components of high-density (HDL) and very low density (VLDL) lipoproteins. Thus, understanding apolipoproteins is very important for medical and health-related fields, such as medical biotechnology, as well as food science and human nutrition.

### 7.1.3. Methods of Structure Determination applied to Apolipoprotein Molecules

The process of biosynthesis, the physical characteristics and the metabolism of apolipoproteins have been intensely studied. However, because of the noncrystalline structure of many apolipoproteins, it has been difficult to obtain structural data at the molecular, or atomic, level. Therefore, methods combining the amino acid sequence with molecular methods are now being introduced. Thus, overall structures may be derived from correlations of global secondary structures determined from polarized light studies combined with local structures predicted from amino acid sequences. The known amino acid sequence of apolipoprotein from the position 7 to 156 of apo Lp-III was first used to design an Apo A-I template, that could be then approached by 'standard' molecular modeling techniques.

### 7.1.4. Molecular Modeling of Apolipoproteins A-I

<u>Target</u>: Apolipoprophorin III is chosen as a target for apolipoprotein A-I.(apo Lp A-I) . The structure of Apolipophorin III has been determined in a crystal at 2.5 Å resolution for the 18-kDa apo Lp III from the African migratory locust, *Locusta migratoria* and the 22-kDa N-terminal, receptor-binding domain of human apo E.

<u>Template for Apo A-I</u>: Lp IIIa is designed by using molecular software IALIGN from Lp III by inserting alanine for template of sequences of apo A-I ( using the program SYBYL, V5.5).  Alanine residues are inserted at each of the gap position identified by IALIGN, an interactive alignment program distributed with the Protein Identification Resource (PIR).(Eleanor M.B, 1994)  This model was compared with DgA-I, HuA-I and ChA-I resprsenting canine, human and chicken Apo A-I respectively.   Results were then compared using a "strip of  the helix" template (Vazquez et al 1992) by scoring 1 or 0 for residues that did , or did not,  fit into the template.

<u>Modeling Results.</u>

   **A**. *Sequence Comparison*.   The five long α-helices connected by short loops in amino acid residues 7-156 of apo Lp-III is used as template for LpIIIa, Dig A-I, DgA-I, HuA-I.  Amphipathic potential (AP) is used to detect if the predict structure is suitable for a lipid-aqueous interface in its stable condition.  The results are shown in **Table 5**.

**Table 5: Amphipathic potentials of predicted helical segments in apolipoprotein models.**

| Model | | | | | | | | | | |
|---|---|---|---|---|---|---|---|---|---|---|
|  | # res | AP | # res | AP | # res | AP | # res | AP | # res | AP |
| Lp-III | 19 | 0.79 | 25 | 0.8 | 22 | 0.77 | 27 | 0.7 | 21 | 0.67 |
| Lp-IIIa | 19 | 0.79 | 25 | 0.8 | 24 | 0.67 | 33 | 0.61 | 22 | 0.64 |
| DgA-I | 19 | 0.58 | 25 | 0.84 | 24 | 0.75 | 33 | 0.48 | 22 | 0.68 |
| HuA-I | 19 | 0.53 | 25 | 0.88 | 24 | 0.71 | 33 | 0.45 | 22 | 0.73 |
| ChA-I | 19 | 0.53 | 25 | 0.88 | 24 | 0.58 | 33 | 0.52 | 22 | 0.82 |

   Amphipathic potentials (AP) are the best average value for a helical segment of  ( # res) residues measured with the "strip of helix".(Vazquez, et al. 1992)

**B.** *Energy minimized models.* In this model, electrostatic interactions contribute the most to favorable energies. Alanine was chosen as the spacer residue in building apo Lp-IIIa because of its function of small, non-ionic side chain and serves as a helix- stabilizing residue. Although it has a high probability of being found in helical structures, it does not participate through electrostatic interactions. Results of this model are discussed for potential energetic evaluation and amphipathic analysis of energy refined helices (see **Table 6 and Table 7**). The structure obtained through this molecular modeling is illustrated in **Figure 6**.

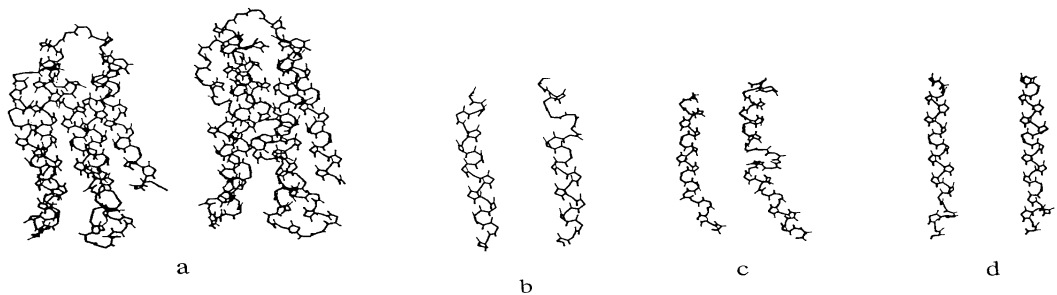

**Figure 6.** Energy Refined models of apo-Lp III and the template apo IIIa constructed by inserting alanine residues into gap positions identified when the sequences were aligned with the canine, human, and chicken apo A-1 sequences. Backbone structures for these complete models are in panel (a). The effects of inserted alanine residues on H3, H4, H5 are displayed in panels (b), (c), (d) respectively. In each display, the apo Lp III is shown to the left of the apo Lp-IIIa. (SOURCE: Brown et al 1994).

**Table 6.   Energetic evaluation of the refined models**

| | Lp-in | Lp-lUa | DgbjA-I | HuA-I | ChA-I |
|---|---|---|---|---|---|
| Number of residues | 150 | 165 | 165 | 165 | 165 |
| Energy, kcal | | | | | |
|    Bond stretching | 20.7 | 23.1 | 28.9 | 29.2 | 31.8 |
|    Angle bending | 123.4 | 154.0 | 189.6 | 208.5 | 212.2 |
|    Torsional | 198.1 | 235.7 | 308.3 | 324.6 | 340.9 |
|    Out of plane | 26.1 | 35.8 | 37.7 | 41.5 | 44.5 |
|    1-4 van der Waals | 218.7 | 235.6 | 247.2 | 260.2 | 270.3 |
|    Van der Waals | -993.0 | -1094.3 | -1074.8 | -1131.8 | -1155.6 |
|    1-4 electrostatics | 1684.0 | 1813.3 | 1497.1 | 1566.7 | 1524.1 |
|    Electrostatic | -4497.8 | -4567.3 | -5214.7 | -5182.6 | -5208.4 |
|    H-bond | -69.3 | -70.2 | -64.5 | -69.8 | -76.3 |
| Total Energy, kcal | -3289.0 | -3234.3 | -4045.2 | -3953.6 | -4016.4 |
|    Kcal/mol/residue | -21.9 | -19.6 | -24.5 | -24.0 | -24.3 |
| H1 (residues) | (7-25) | | (72-90) | (73-88) | (72-90) |
|    kcal/mol/residue | -16.1 | | -18.5 | -16.6 | -17.4 |
| H2 (residues) | (35-59) | | (100-124) | (101-125) | (100-124) |
|    kcal/mol/residue | -16.1 | | -20.1 | -19.4 | -19.2 |
| H3 (residues) | (70-91) | (70-91) | (135-158) | (136-159) | (135-158) |
|    kcal/mol/residue | -17.6 | -20.7 | -18.4 | -16.2 | -17.6 |
| H4 (residues) | (96-121) | (96-121) | (166-196) | (167-197) | (166-196) |
|    kcal/mol/residue | -15.5 | -13.8 | -16.7 | -16.9 | -17.6 |
| H5 (residues) | (136-155) | (136-155) | (207-235) | (208-236) | (207-235) |
|    kcal/mol/residue | -15.8 | -15.4 | -14.0 | -13.4 | -14.9 |

These energy calculations, based on the sequence segments initially assigned to HI -H5 (see Tables 6 and 7), are given to illustrate the stability of the structures. Sequences for helices 3 to 5  (H3, H4, H5) of apo Lp-IIIa contain inserted Alanine residues. (SOURCE: Brown 1994).

**Table 7. Amphipathic analysis of energy refined helices**

| Model | Lp-III | Lp-IIIa | gA-I | HuA-I | ChA-I |
|---|---|---|---|---|---|
| H1 | N7-E25 | N7-E25 | D72-E90 | D73-K88 | E72-E90 |
| Hb sector | 140° | 140° | 100° | 60° | 100° |
| Av. Hb | 0.98 | 0.98 | 0.78 | 0.7 | 0.81 |
| H2 | P35-S59 | P35-S59 | L100-E124 | L101-E125 | L101-E124 |
| Hb sector | 100° | 100° | 100° | 100° | 100° |
| Av. Hb | 0.8 | 0.8 | 0.7 | 0.89 | 0.95 |
| H3 | S70-T91 | S70-S85 | Q137-R150 | E136-R149 | L135-L151 |
| Hb sector | 140° | 120° | 100° | 100° | 120° |
| Av. Hb | 0.93 | 0.79 | 0.95 | 0.88 | 0.82 |
| H4 | | | | | |
| H4A | A96-S121 | Q98-T107 | D168-K181 | Q172-E179 | D167-R181 |
| Hb sector | 100° | 140° | 80° | 100° | 80° |
| Av. Hb | 0.56 | 0.91 | 1.1 | 0.94 | 1.18 |
| H4B | _______ | Q117-al21d | S187-S196 | A190-A194 | P186-V195 |
| Hb sector | | 60° | 100° | _______ | 60° |
| Av. Hb | _______ | 0.45 | 0.45 | | 0.66 |
| H5 | E136-V156 | L134-A155 | L213-I232 | L214-T237 | E212-L235 |
| Hb sector | 120° | 80° | 100° | 100° | 100° |
| Av. Hb | 0.74 | 0.53 | 1.05 | 0.96 | 0.86 |

**Table 7.** The size of the hydrophobic (Hb) sector of a helix was determined from helical Wheel projections (Schiffer et al 1967) The average hydrophobicity (Av. Hb), of this sector was calculated by the method of Eisenberg (1984). Single letter designations are used for Amino acid residues, (al21d) designates the fourth Alanine residue inserted between residues 121 and 122 of Lp-III. (SOURCE: Brown 1994).

After evaluating the potential energy for this model, the lateral view structures of apolipoproteins for apo lipophorin III(residues 7-156), canine apolipoptrotein A-I (residues 72-236), human apolipoprotein A-I (residues 73-237) and chicken apolipoprotein A-I (residues 72-236). The computation results are shown in **Figure 7** (according to Brown 1994).

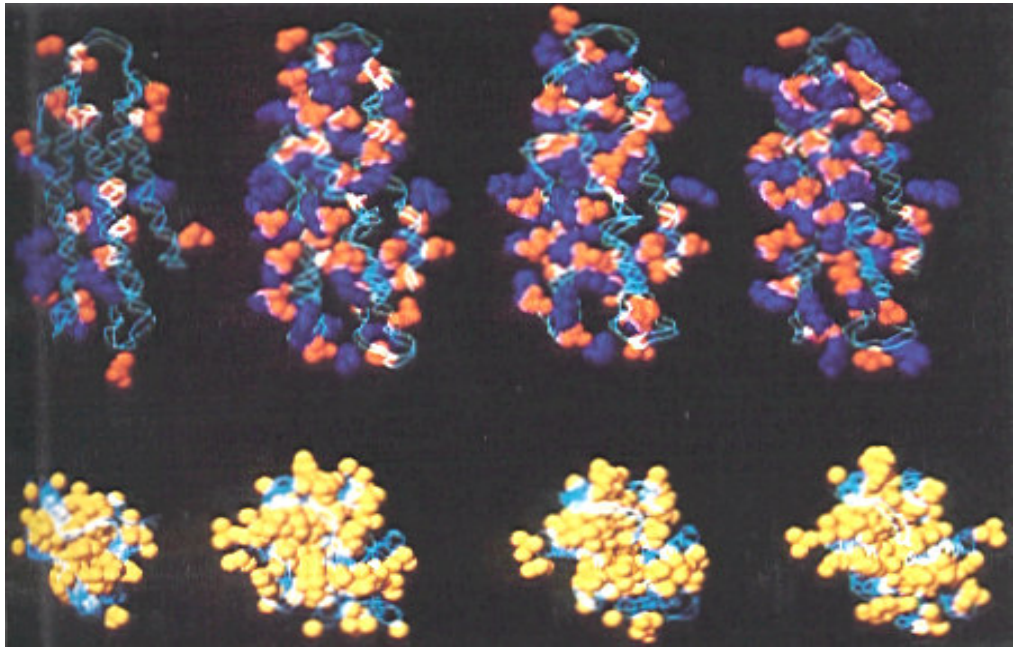

**Figure 7.** Energy refined models of from left to right apo lipophorin III(residues 7-156) canine apolipoptrotein A-I (residues 72-236), human apolipoprotein A-I (residues 73-237) and chicken apolipoprotein A-I (residues 72-236). Peptide backbones are represented by a double stranded ribbon. In the lateral view, only ionizable side-chains are displayed with acidic groups in red and basic groups in blue. In the end-on view of the molecules only the hydrophobic side-chains are displayed in orange. (SOURCE: Brown 1994)

## 7.2.    *Combination of Molecular Modeling with Other Techniques*

It is important to combine molecular modeling with other techniques in order to improve the accuracy of the modeling results. A recent study of Apolipoprotein has produced a high-resolution reconstruction of the structure of apolipoprotein through combination with a solution-phase X-ray technique. It was shown that Apoliproprotein A-1 is a 243-residue protein that contains a globular amino-terminal domain (residue 1-43) and a lipid-binding carboxyl-terminal domain (residues 44-243; Segrest et al 1992). The aqueous phase X-ray crystal structure was obtained at 4 Å resolution. This has suggested the accuracy of molecular modeling by combining the X-ray crystallographic with molecular modeling (Figure 8; from Segrest et al 1992; Borhani 1997).

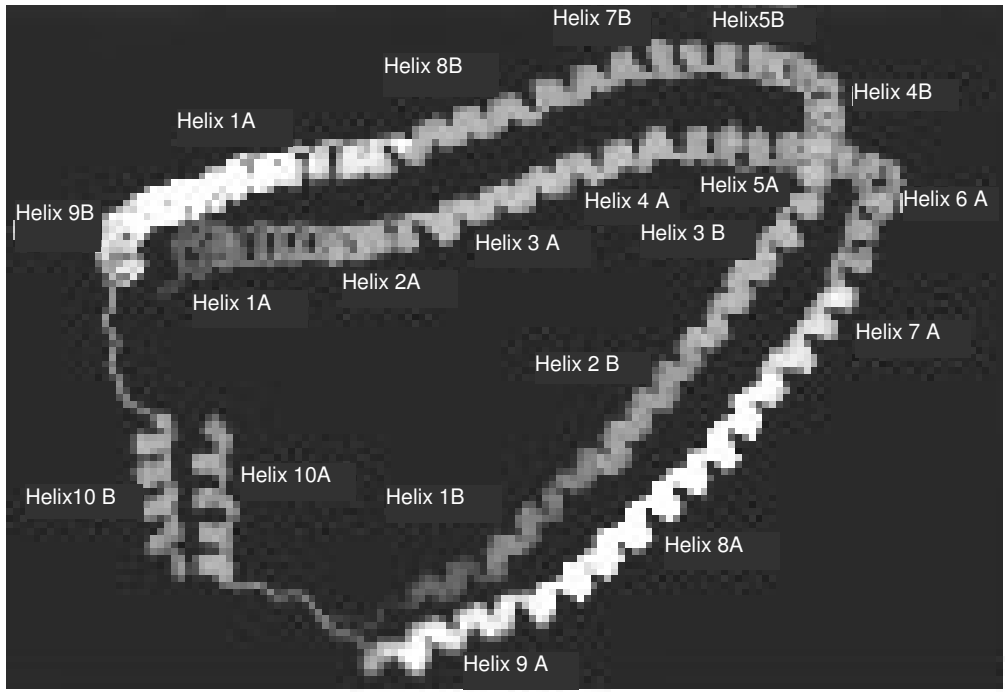

**Figure 8.** X-ray Crystal structure of the Apolipoprotein A-I: Δ( 1-43) dimer in solution.
(Source:  Borhani et al.  1997).

### 7.3.  An In-depth Analysis of Molecular Structure

The determination of apolipoprotein A-I can be then further associated with lipid containing domains by employing other molecular modeling techniques. The 'belt model' is used to show the possible orientations of lipoprotein with its apolipoprotein inserted (**Figure 9**). The suggested structure can then serve as a template in other high density lipoproteins for their structure determination and also help in understanding the biological interaction.

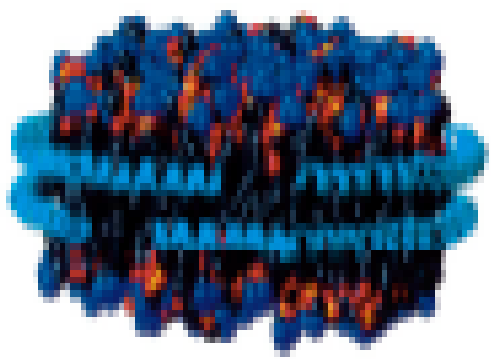
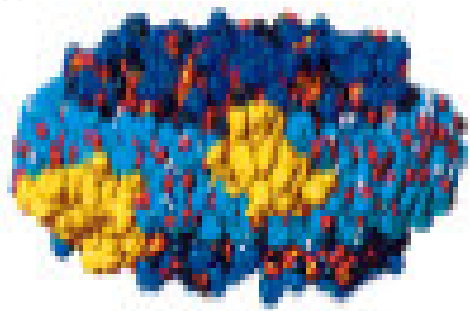

**Fig 9a.** Detailed Belt model displayed as a helical ribbon. C(NH$_2$)$_3$, blue, oxygen atom, red; phosphorous atom, yellow all other atom, black. ( Source: Borhani 1997)

**Fig 9b.** Detailed Belt model displayed as all atom model, oriented in 9a. nitrogen atom, blue, oxygen atom, red; carbon atom, cyan, polar hydrogen atom, white. ( Source: Borhani 1997)

### 7.4. Applications of Molecular Modeling

Molecular modeling has been introduced for more than two decades ago. Increasingly, modeling software is available for a variety of industrial applications. Global markets for molecular modeling, in general, now exceed 2 billion US $ annually (Fuji-Keizai 2003)

### 7.4.1. Applications in the Food Industry and Human Nutrition

Molecular modeling has been suggested by several professionals in food industry as a new tool for food research. (Hegenbart 1992)  Such tools can assist food scientists in problem solving, as well as save time and money. Examples of utilization of molecular modeling are the uses of high intensity sweeteners and taste receptors to predict the sweetening potential of new molecules by using molecular modeling as developed by and E.W. Taylor and S. Wilson at the University of Georgia. (Hegenbart 1992)  Such models can be used in food industry for product development and also for faster results in sensory evaluation.

### 7.4.2. Examples of Medical Applications of Molecular Modeling

Molecular modeling is especially helpful in medical fields, such as in development of new drugs on a nanoscale. Recent studies have shown the importance of using molecular modeling in both medical and food sciences (Food Ingredient first, 2003). The molecular modeling of Epigallocatechin Gallate (EGCG), and the HIV cell was undertaken by Shearer (2003). His report has inspired scientists in Japan who discovered the potential of green tea as an anti-HIV drug. The chemical compound that is found abundantly in the green tea called *Epigallocatechin Gallate* (EGCG) is reported to stop the HIV virus from binding to CD4 molecules and human T-cells.

### 7.4.3. Other Applications of Molecular Modeling

Other applications of molecular modeling to manufacturing, life sciences and chemistry greatly benefit from such molecular modeling programs. Nanotechnology has developed to a 30 to 40 million US $ market, and it also has the potential to grow to a 60 to 70 million US $ market within the next five years (Fuji-Keizai 2003).

## 8. CONCLUSIONS

A simplified overview of the selected applications of biotechnology in the areas of foods, human nutrition and health, as well as the potential, large-scale applications in the chemical industry that were discussed in our review is presented in **Table 8**.

| NOVEL TECHNIQUES | AREA | APPLICATION |
|---|---|---|
| Immobilized cells | Food Applications | Brewing industry |
| | | Dairy Foods |
| | | |
| | Human Nutition | Neutraceuticals |
| | | |
| | Medical Applications | Regulation of fibrinolisis |
| | | Detoxification of water |
| Genetic Manipulation | Food Applications | Synthesis of complex carbohydrates |
| | | cloning of cellulose degrading enzyme |
| | Human Nutrition | Addition of vitamins  in foods |
| | | Reduction of ripening |
| | Medical applications | |
| | Other | Bioproduction of fuel |
| Liposomes in Microencapsulation | Food Applications | Flavor delivery |
| | | Additives in bakery |
| | | Chesee ripening |
| | Medical applications | protecion of DNA interactions |
| | | Gene therapy |
| | | Vaccines |
| | | Drug carrier against Cancer and HIV |
| | Other | Creams |
| | | Oil spills separations |

Table 8. Selected Biotechnology Tools and Their Applications

The rapid development of biotechnological tools and 'quick' applications oriented towards immediate marketing may be responsible for generating legislative barriers against 'GMO-based' products. The complementary area of genetically manipulated microorganisms is adopting novel approaches to overcome such increasingly unprofitable legislative barriers and boost profits both in the short- and long-term.

The use of ultra-sensitive techniques and bio-computational modeling is essential in order to quantitate the physical and chemical properties of molecules and supra-molecular systems that are of primary interest to developments in Biotechnology and its applications.

Policy changes may be therefore considered and implemented that would take advantage of such novel approaches to develop new niches and markets for both profitable and safe Biotechnology applications worldwide, in the Chemical Industry, Agriculture, Health and Medicine.